\documentclass[12pt]{iopart}
\usepackage{iopams}
\usepackage{epsfig}   
\usepackage{graphics}  
\begin{document}

\title[Tentative detection of the gravitational 
magnification of type Ia SNe]{Tentative detection of the gravitational
magnification of type Ia supernovae}

\author{Jakob J\"onsson$^1$, Tomas Dahl\'en$^2$, Ariel Goobar$^1$, 
  Edvard M\"ortsell$^3$ and Adam Riess$^{2,4}$}

\address{$^1$ Department of Physics, Stockholm University, Albanova
         University Center \\
         S--106 91 Stockholm, Sweden}
\address{$^2$ Space Telescope Science Institute, 3700 San Martin Drive,
  Baltimore, MD 21218, USA} 
\address{$^3$ Department of Astronomy, Stockholm University, Albanova
         University Center \\
         S--106 91 Stockholm, Sweden}
\address{$^4$ Johns Hopkins University, 3400 North Charles Street, 
  Baltimore, MD 21218}

\ead{\mailto{jacke@physto.se}}

\begin{abstract}
The flux from distant type Ia supernovae (SN) is likely to be
amplified or de-amplified by gravitational lensing due to matter
distributions along the line-of-sight. A gravitationally lensed SN
would appear brighter or fainter than the average SN at a particular
redshift.  We estimate the magnification of 26 SNe in the GOODS fields
and search for a correlation with the residual
magnitudes of the SNe. The residual magnitude, i.e.~the difference
between observed and average magnitude predicted by the ``concordance
model'' of the Universe, indicates the deviation in flux from the
average SN.  The linear correlation coefficient for this sample is
$r=0.29$. For a similar, but uncorrelated sample, the probability of
obtaining a correlation coefficient equal to or higher than this value
is $\sim 10\%$, i.e. a tentative detection of lensing at $\sim 90\%$
confidence level.
Although the evidence for a correlation is weak, our result is in
accordance with what could be expected given the small size of the
sample.
\end{abstract}

\noindent{\it Keywords}: gravitational lensing, supernova type Ia

\section{Introduction}
Matter inhomogeneities in the Universe, such as galaxies, groups, and
clusters of galaxies, give rise to gravitational fields which affect
the path of light rays emitted from distant sources.  Gravitational
bending of light, so called gravitational lensing, can amplify or
de-amplify the flux from distant sources. A gravitationally lensed
standard candle, such as a type Ia supernovae (SN), at a particular
redshift would thus appear brighter or fainter to an observer than the
average standard candle at this redshift.  Using observations of the
matter in the foreground, we can estimate the magnification factor
$\mu$, which describes the amplification ($\mu > 1$) or
de-amplification ($\mu < 1$) relative to a homogeneous universe, of a
line-of-sight. The amplified flux of a point source with flux $f$ is
given by $\mu f$.  If our estimates of the amplification are correct,
we expect to find a correlation between standard candle brightness and
magnification.  In this letter we use SNe to search for this
correlation.
Gravitational lensing adds extra scatter, which is 
expected to grow with redshift 
\cite{hol05}, to the
intrinsic dispersion of SN brightness.  However, most of this
dispersion is due to a few highly magnified SNe and a large portion of
the SNe are slightly de-magnified.  The effects of intrinsic scatter
and measurement errors are thus larger than the effect of
gravitational lensing for most SNe and the correlation might
consequently be hard to reveal.

The correlation between foreground matter and SN brightness
has been searched for earlier.
Williams and Song \cite{wil04} reported a correlation between
luminosity of SNe and the density of foreground galaxies
within $5-15$ arcminutes radius from the position of the SN.
M\'enard and Dalal \cite{men05} searched for a correlation on scales
$1-10$ arcminutes, but could not find any correlation. 
In this letter we search for a correlation between
SN brightness and the magnification due to foreground galaxies
within $\sim 1$ arcminute. 
Since we use spectroscopic or photometric 
redshifts of the galaxies, we take the spatial
distribution of matter into account. The possible confusion
of foreground and background galaxies have been incorporated in our
error estimation.

\section{Data}
In order to reconstruct the amplification along a
line-of-sight we need galaxy positions, redshifts, and luminosities.
Hubble Deep Field North (HDFN) and Chandra Deep Field South
(CDFS) observed within the Great Observatories Origins Deep Survey
(GOODS) are ideal for our purposes, since they have been extensively
imaged in many wavelength bands \cite{cap04,gia04} and have been 
extensively searched for SNe \cite{str04,rie04,rie07}.
We use the information from the GOODS to estimate the magnification of
$26$ SNe observed in these fields.
Since most of the galaxies lack spectroscopic redshifts 
(especially high redshift galaxies)
we have to rely on photometric redshifts which depend on observations in
many different wavelength bands.
The galaxy data we have used have been described in detail in Section 3 in 
\cite{jon06}.

\section{Estimating the magnification factor}
To compute the magnification factor of the SNe we have used the
prescription described in \cite{jon06}. Since each galaxy along the
line-of-sight can influence the path of a light ray,  we use
a numerical code (QLET)\cite{gun04} utilizing the multiple lens
plane formalism and include all foreground galaxies within $1$
arcminute from the
SN in the calculations. Galaxies are modelled as truncated Singular
Isothermal Spheres (SIS) or truncated Navarro, Frenk, and White
profiles (NFW)\cite{nav97}. The mass of the galaxy halos were
estimated using the luminosities of the galaxies via empirical
Tully-Fisher and Faber-Jackson relations. The discrepancy between
the total mass of the galaxies in the fields and the total amount of
matter ($\Omega_{\rm M}\sim 0.3$) in the Universe, is distributed 
homogeneously.
Errors in the magnification factor of each SN were estimated
using Monte-Carlo simulations, taking into account photometric
redshift uncertainties and scatter in the Tully-Fisher and
Faber-Jackson relations, following the method outlined in Section 4 
in \cite{jon06}. However, the largest uncertainty is due to the
uncertainty in the halo model, which is reflected by the differences
in $\mu$ calculated using SIS or NFW halo profiles. 
We have estimated the magnification factor for $26$ SNe in the GOODS 
fields for which the distance modulus have been accurately 
measured\footnote{We have only considered ``gold'' SNe.} \cite{rie07}. 
For 14 of these SNe magnification
factors were presented already in \cite{jon06}. 
Magnification factors in logarithmic units, 
$\Delta_{\rm \mu}=-2.5\log_{10}\mu$, estimated for each SN assuming
galaxy halos to be described by NFW- or SIS-profiles are listed in Table
\ref{tab:dat}.

\Table{\label{tab:dat} Residual magnitudes and magnification of supernovae.}
\br
&&\centre{2}{Magnification$^{\rm a}$ (mag)}\\
\ns
&&\crule{2}\\
Supernova & Residual (mag) & NFW & SIS\\
\mr
 1997ff    &$        -0.48  \pm    0.35  $&$       -0.177  \0   \pm   0.052  
 $&$       -0.130   \0  \pm   0.040  $\\
 2002fw    &$\m       0.04  \pm    0.20  $&$\m      0.027  \0   \pm   0.017  
 $&$\m      0.022   \0  \pm   0.016  $\\
 2002hp    &$        -0.52  \pm    0.30  $&$\m      0.006  \0   \pm   0.025  
 $&$\m      0.010   \0  \pm   0.020  $\\
 2002kd    &$        -0.34  \pm    0.19  $&$\m      0.001  \0   \pm   0.017  
 $&$\m      0.002   \0  \pm   0.013  $\\
 2002ki    &$\m       0.05  \pm    0.29  $&$\m      0.053  \0   \pm   0.010  
 $&$\m     0.0539       \pm  0.0094  $\\
 2003bd    &$        -0.05  \pm    0.24  $&$\m      0.006  \0   \pm   0.011  
 $&$\m     0.0035       \pm  0.0089  $\\
 2003be    &$        -0.10  \pm    0.25  $&$      -0.0018       \pm  0.0094  
 $&$\m     0.0039       \pm  0.0077  $\\
 2003dy    &$        -0.17  \pm    0.31  $&$\m      0.017  \0   \pm   0.018  
 $&$\m      0.017   \0  \pm   0.015  $\\
 2003eb    &$        -0.39  \pm    0.25  $&$\m      0.031  \0   \pm   0.010  
 $&$\m     0.0235       \pm  0.0094  $\\
 2003eq    &$        -0.17  \pm    0.21  $&$\m     0.0398       \pm  0.0089  
 $&$\m     0.0355       \pm  0.0081  $\\
 2003es   &$\m       0.12  \pm    0.27  $&$       -0.035  \0   \pm   0.014  
 $&$       -0.016   \0  \pm   0.012  $\\
 2003lv    &$        -0.16  \pm    0.20  $&$\m      0.054  \0   \pm   0.013  
 $&$\m      0.048   \0  \pm   0.013  $\\
 HST04Eag  &$\m       0.16  \pm    0.19  $&$       -0.075  \0   \pm   0.043  
 $&$       -0.050   \0  \pm   0.035  $\\
 HST04Gre  &$        -0.22  \pm    0.31  $&$       -0.061  \0   \pm   0.051  
 $&$       -0.043   \0  \pm   0.039  $\\
 HST04Man  &$\m       0.07  \pm    0.29  $&$\m     0.0440       \pm  0.0073  
 $&$\m     0.0433       \pm  0.0071  $\\
 HST04Mcg  &$\m       0.07  \pm    0.25  $&$\m      0.040  \0   \pm   0.026  
 $&$\m      0.039   \0  \pm   0.022  $\\
 HST04Omb  &$        -0.03  \pm    0.26  $&$\m     0.0387       \pm  0.0092  
 $&$\m     0.0358       \pm  0.0079  $\\
 HST04Rak  &$        -0.12  \pm    0.22  $&$\m      0.006  \0   \pm   0.018  
 $&$\m      0.010   \0  \pm   0.014  $\\
 HST04Sas  &$        -0.29  \pm    0.19  $&$       -0.029  \0   \pm   0.032  
 $&$       -0.019   \0  \pm   0.023  $\\
 HST04Tha  &$        -0.33  \pm    0.27  $&$       -0.024  \0   \pm   0.019  
 $&$       -0.021   \0  \pm   0.016  $\\
 HST04Yow  &$        -0.01  \pm    0.32  $&$\m     0.0181       \pm  0.0029  
 $&$\m     0.0180       \pm  0.0022  $\\
 HST05Fer  &$        -0.37  \pm    0.27  $&$       -0.009  \0   \pm   0.015  
 $&$       -0.007   \0  \pm   0.014  $\\
 HST05Gab  &$\m       0.05  \pm    0.18  $&$\m      0.041  \0   \pm   0.015  
 $&$\m      0.038   \0  \pm   0.013  $\\
 HST05Lan  &$\m       0.11  \pm    0.20  $&$       -0.088  \0   \pm   0.046  
 $&$       -0.063   \0  \pm   0.034  $\\
 HST05Spo  &$        -0.39  \pm    0.20  $&$\m      0.006  \0   \pm   0.016  
 $&$\m      0.005   \0  \pm   0.013  $\\
 HST05Str  &$\m       0.43  \pm    0.19  $&$\m     0.0573       \pm  0.0078  
 $&$\m     0.0542       \pm  0.0081  $\\

\br
\end{tabular}
\item[] $^{\rm a}$ Quoted errors are root mean square deviations from
  the mean.
\end{indented}
\end{table}

\section{The correlation}
Since flux is conserved, the average flux from a large number of
lensed standard candles, all at the same distance from the observer, 
is expected to be the same as the flux from a standard candle in an
homogeneous universe. The average brightness or distance modulus of a
standard candle can thus be predicted from a cosmological model of a
homogeneous universe. We take the ``concordance model'', a flat
universe with a cosmological constant and matter density $\Omega_{\rm
  M}=0.29$, to be our model of the Universe. 
In Table \ref{tab:dat} residual magnitudes, $\Delta$, of the SNe 
calculated by subtracting the
predicted distance moduli from the observed value are presented. SNe brighter
than average thus have negative residuals, while fainter than average SNe have
positive residuals.

\begin{figure}
\begin{center}
\includegraphics[angle=-90,width=1.\textwidth]{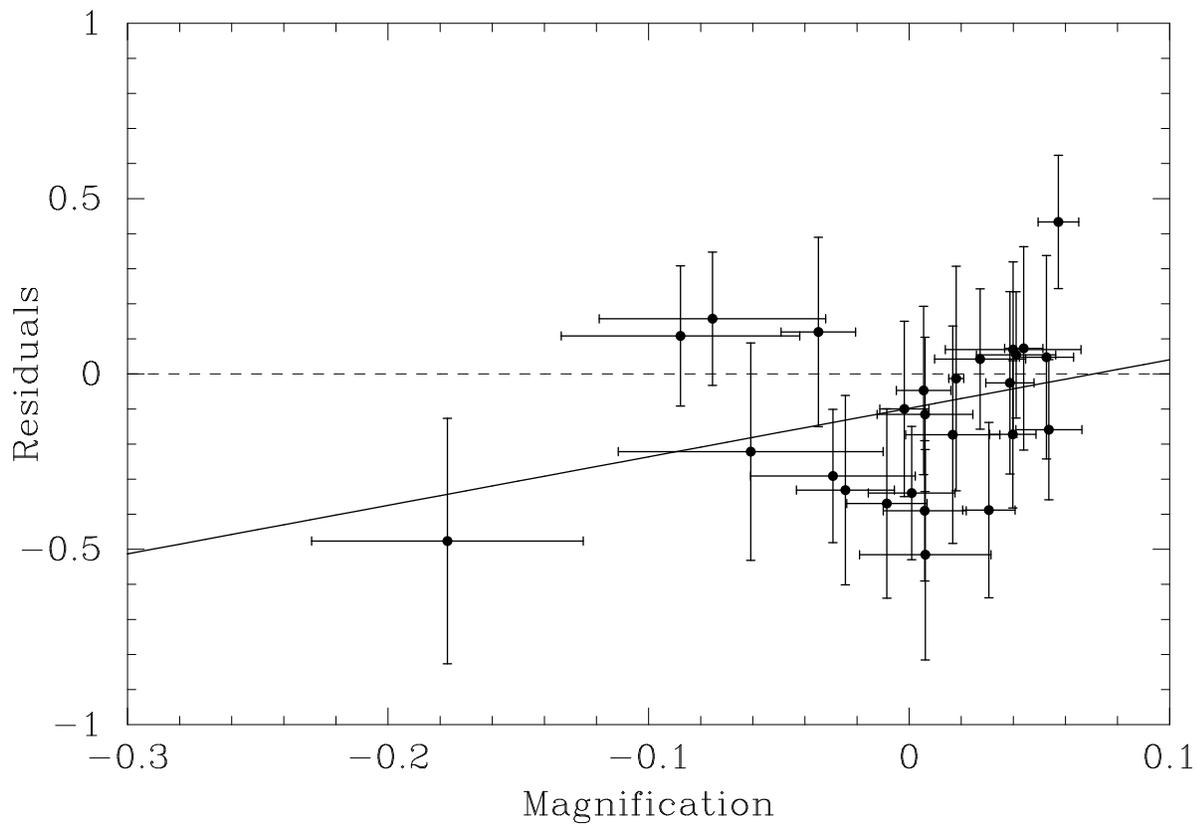}
\caption{\label{fig:corr} Residuals vs magnifications.
  The solid line shows the best fit straight line to the data. The dashed
  line corresponds to the case of no correlation.}
\end{center}
\end{figure}

Figure \ref{fig:corr} shows a scatter plot of residuals vs
magnifications, the latter
computed using NFW-profiles and expressed in logarithmic units 
$\Delta_{\rm \mu}=-2.5\log_{10}\mu$. For the sample considered here the
linear correlation coefficient is $r=0.29$ (the correlation
coefficient is slightly larger, $r=0.34$,  if the magnification
is computed using SIS-profiles).
If we instead compute a 
rank-order correlation coefficient (Spearman), we get a slightly
lower value $r_{\rm s}=0.27$ ($r_{\rm s}=0.31$ for the SIS case). 
For a linear correlation between magnification and residuals, the data
should give a reasonable fit to a straight line, 
$\Delta =a+b\Delta_{\mu}$. Errors are large in both
residuals and magnifications, therefore we take into account errors in both
coordinates\footnote{The best fit straight line is found by minimizing
  \begin{displaymath} \chi^2=\sum_i\frac{
  \left(\Delta_i-a-b\Delta_{{\mu}i}\right)^2}
    {\sigma_{{\Delta}i}^2+b^2\sigma_{\Delta_{\mu}i}^2}, 
  \end{displaymath}
  with respect to $a$ and $b$. Uncertainties in residuals and
  magnifications are denoted by $\sigma_{\Delta i}$ and 
  $\sigma_{\Delta_{\mu} i}$, respectively.}. 
The best fit to the data is indicated
by the solid line in Figure \ref{fig:corr}. For this line ($a=-0.098
\pm 0.048$ and $b=1.4 \pm 1.4$) the
chi-square is $\chi^2=24.09$ for $24$ degrees of freedom.  If there is
no correlation we expect a line with $a=0$ and $b=0$, indicated by the
dashed line in the figure, to be a good fit to the data. The
chi-square for this line is $\chi^2=29.24$, indicating a weak
preference ($\sim 90\%$ confidence level) for a positive correlation.

%

\section{Discussion}
\begin{figure}
\begin{center}
\includegraphics[angle=-90,width=1.\textwidth]{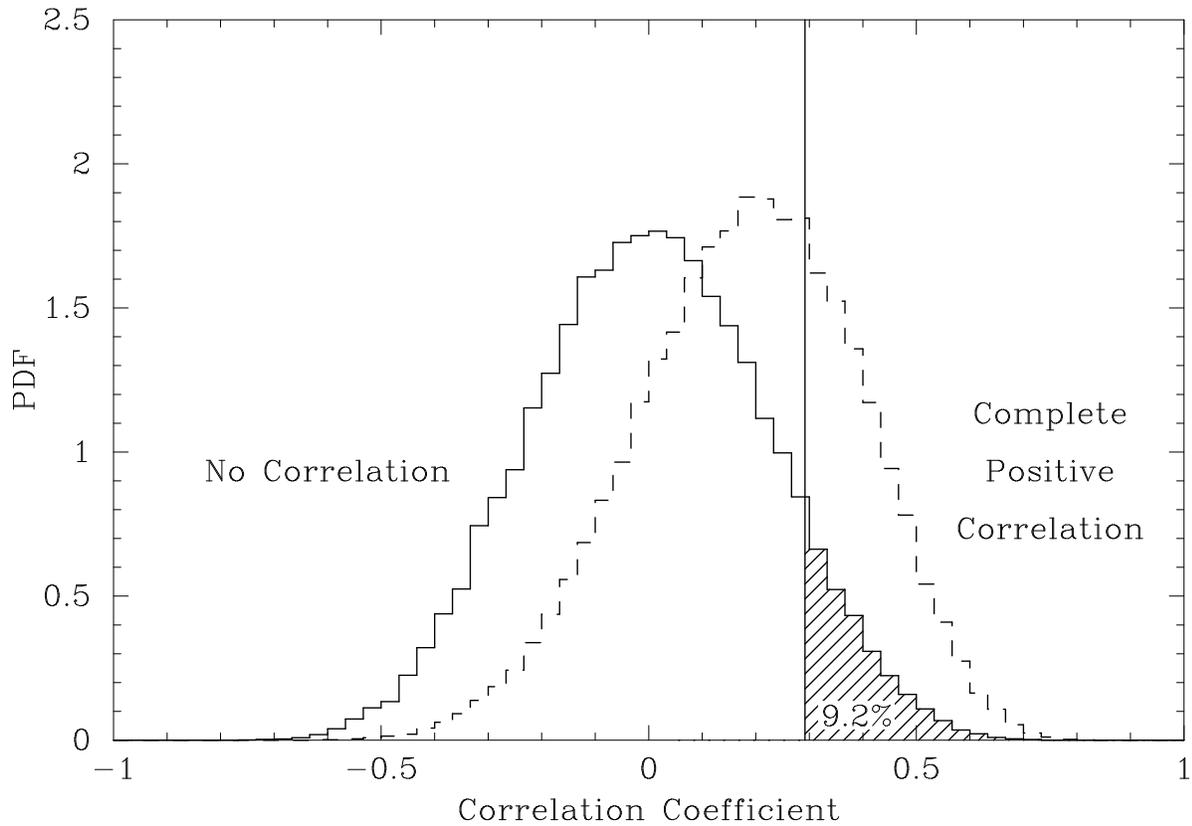}
\caption{\label{fig:mc} Probability distribution functions (PDF) of linear
  correlation coefficient for simulated data sets. The solid curve
  shows the distribution obtained for uncorrelated simulated data. The
  dashed curve shows the distribution obtained for simulated data with
  a complete positive correlation between magnifications and
  residuals. The value of the linear correlation coefficient
  obtained for the data ($r=0.29$) is indicated by the vertical solid
  line. 
  The probability of obtaining a correlation coefficient higher than this
  value in the case of no correlation is $9.2\%$. }
\end{center}
\end{figure}

The evidence for a correlation between residuals and magnifications is
rather weak. Considering the low statistics, what could we
expect to measure? Even if there exists a complete positive
correlation between magnifications and residuals, the intrinsic
dispersion in SN luminosity and measurement errors would hide the
correlation. The strength of the correlation we can expect to measure
depend on the size and quality of the data set. 
To examine the strength of the correlation we have made the following 
simple experiment. We assumed a complete
positive correlation between magnification and residuals. The residual
of a SN, the magnification of which is $\mu$, would then be given by
$\Delta =\Delta_{\mu}$.
To this residual we added normally distributed
noise according to the measurement error of the SN magnitude
and an error in the cosmological model used to predict the average 
distance modulus which was subtracted from the observed value. 
The error in the cosmological model corresponds to $10\%$ 
uncertainty in $\Omega_{\rm M}$. 
This noise hides the otherwise complete correlation. 
We repeated this experiment several times and the result is shown as the dashed
curve in Figure \ref{fig:mc}. The most likely linear correlation
coefficient is $r=0.18$.
As can be seen from the figure a large range of correlation
coefficients can be expected for a sample of the size considered here. 
The measured
correlation coefficient ($r=0.29$), indicated by the vertical solid line, is
in accordance with the expected strength of the correlation.

We can also asses the probability of measuring the value of the
correlation coefficient we obtain, if there were no correlation. The
solid curve in Figure \ref{fig:mc} shows the result of computing the
correlation coefficient for a large number of simulated data sets
where SN residuals and magnifications are uncorrelated. The
probability of obtaining a correlation factor of $r=0.29$ or higher is 
$9.2\%$. 
Since the probability of no correlation is $9.2\%$, the confidence
level of our detection of the correlation is $90.8\%$.
In the case where the magnification was computed using
SIS-profiles the probability of obtaining a correlation factor higher
or equal to the measured value ($r=0.34$) is
$5.8\%$.

If we assume the uncertainty in our ``concordance model'' to be
$10\%$ in $\Omega_{\rm M}$, the resulting distribution of linear
correlation coefficients would be a normal distribution with a
standard deviation of $0.01$ (irrespective of the halo model used). The
probability of finding a
correlation coefficient 3 standard deviations higher ($r=0.32$) or 
lower ($r=0.26$) than the
measured value ($r=0.29$) for an uncorrelated data set is $7.2\%$ and 
$12.0\%$, respectively. Our results are thus not very sensitive to
errors in the cosmology used to predict the average SN brightness.

The evidence we find for a correlation between SN residuals and
magnifications are weak and should be regarded as tentative at the
best. However, the effect should be detectable at higher confidence
given a larger data set. To find a correlation of $r=0.18$ (the most
likely value if we model galaxies as NFW-profiles) at the $99\%$
confidence level requires $220$ SNe with the same distribution of
redshift ($0.4 \lesssim z \lesssim 1.8$) and similar uncertainties as
the small sample considered here.

Trustworthy evidence of a correlation would indicate that our models
of the matter inhomogeneities in the Universe are reliable and would
explain some of the scatter in the distance-redshift relation. This
scatter could then be reduced, at least partially, by correcting for
the magnification of individual SNe as suggested by \cite{gun06}.
%

\ack
AG would like to acknowledge support by the Swedish Research Council
and the G\"oran Gustafsson Foundation for Research in Natural Sciences
and Medicine. EM acknowledge support for this study by the Swedish
Research Council and from the Anna-Greta and Holger Crafoord fund.


\end{document}